# Investigating Student Difficulties with Dirac Notation

Chandralekha Singh and Emily Marshman

*Department of Physics and Astronomy, University of Pittsburgh, Pittsburgh, PA, 15260, USA*

**Abstract**: Quantum mechanics is challenging even for advanced undergraduate and graduate students. Dirac notation is a convenient notation used extensively in quantum mechanics. We have been investigating the difficulties that the advanced undergraduate and graduate students have with Dirac notation. We administered written free response and multiple-choice questions to students and also conducted semi-structured individual interviews with 23 students using a think-aloud protocol to obtain a better understanding of the rationale behind their responses. We find that many students struggle with Dirac notation and they are not consistent in using this notation across various questions in a given test. In particular, whether they answer questions involving Dirac notation correctly or not is context dependent.

## INTRODUCTION

Learning quantum mechanics concepts can be challenging even for advanced undergraduate and graduate students in physics [1-6]. We have been examining student difficulties in learning quantum mechanics. Here, we focus on an investigation to identify students' difficulties with Dirac notation. Because Dirac notation is used so extensively in quantum mechanics, it is important that students have a thorough understanding of this notation. The investigation was conducted with students at the University of Pittsburgh (Pitt) by administering written free response and multiple-choice questions as a part of undergraduate or graduate level courses and by conducting in-depth individual interviews with 23 students using a think-aloud protocol outside of class. Students were asked to talk aloud while they answered questions posed to them. In these semi-structured interviews, the interviewers did their best not to disturb students' thought processes while they answered the questions except to urge them to keep talking if they became quiet for a long time. Later, the interviewer asked students for clarification of points they had not made clear earlier in order to understand their thought processes better. Some of these clarification questions were planned out ahead of time while others were emergent queries based upon a particular student's responses during an interview.

Here, we present some findings related to the difficulties that physics graduate students have with Dirac notation as manifested by their written responses to conceptual multiple choice questions in two different graduate classes. These questions were posed after eleven weeks of the first semester of graduate quantum mechanics for one set of graduate students (15 total number) and almost full two semesters of the graduate quantum mechanics course for the second set of graduate students (24 total number). We found that there is no significant difference between the scores of the two graduate classes on the multiple choice questions. We also discuss the findings from think-aloud interviews conducted with individual graduate students to understand their difficulties in-depth after they had successfully completed a two-semester graduate quantum mechanics course.

## STUDENT DIFFICULTIES

Our investigation suggests that many students' responses are context dependent in that students were inconsistent in their responses to different questions about Dirac notation given as a part of the same test. This indicates that even graduate students do not yet have a good grasp of the notation, e.g., in recognizing position space and momentum space wavefunctions in Dirac notation in different contexts. Below, we discuss 39 physics graduate students' performance (combining the performances from two different years since the performances are similar) on five conceptual multiple choice questions related to Dirac notation. Interviews revealed certain aspects of Dirac notation that were particularly challenging and lead to inconsistencies.

**TABLE 1**. Percentages of graduate students who selected answer choices for the five multiple choice questions about Dirac notation. The percentages may not add to 100% because some students did not answer the question. (The correct answers are italicized.)

|  | A | B | C | D | E |
|---|---|---|---|---|---|
| Question 1 | 0 | 3 | *85* | 5 | 5 |
| Question 2 | 18 | 18 | *51* | 8 | 3 |
| Question 3 | 13 | 13 | 36 | 5 | *31* |
| Question 4 | 13 | 10 | *54* | 8 | 10 |
| Question 5 | 21 | 15 | 13 | 10 | *41* |

Question 1 probes students' grasp of position and momentum space wavefunctions in Dirac notation.

*For a spinless particle confined in one spatial dimension, the state of the quantum system at time t=0 is denoted by $|\psi\rangle$ in the Hilbert space. $|x\rangle$ and $|p\rangle$ are the eigenstates of position and momentum operators. Answer questions 1-4.*

1. *Choose all of the following statements that are correct about the position space and momentum space wavefunctions for this quantum state.*

(1) The position space wavefunction is $\psi(x) = \langle x|\psi\rangle$ where x is a continuous index.

(2) The momentum space wavefunction is $\psi(p) = \langle p|\psi\rangle$ where p is a continuous index.

(3) The momentum space wavefunction is $\psi(p) = \frac{\hbar}{i}\frac{\partial|\psi\rangle}{\partial x}$

A. 1 only   B. 2 only   C. 1 and 2 only   D. 3 only   E. 1 and 3 only

Table 1 shows that students performed well on this question with 85% selecting the correct answer (Choice C). However, student responses to other questions discussed below suggests that many students do not necessarily recognize that $\psi(x) = \langle x|\psi\rangle$ and $\psi(p) = \langle p|\psi\rangle$ in other contexts.

The following multiple choice question investigates difficulties with position space and momentum space wavefunctions written in Dirac notation and how they are related by a Fourier transform.

2. *Choose all of the following equations involving the inner product that are correct.*

(1)   $\langle x|\psi\rangle = \int x\psi(x)dx$

(2)   $\langle x|\psi\rangle = \int \delta(x-x')\psi(x')dx'$

(3)   $\langle p|\psi\rangle = \int \langle p|x\rangle\langle x|\psi\rangle dx = \int e^{-ipx/\hbar}\psi(x)dx$

(4)   $\langle p|\psi\rangle = \int \frac{\hbar}{i}\frac{\partial}{\partial x}\psi(x)dx$

A. 1 and 3 only   B. 1 and 4 only   C. 2 and 3 only   D. 2 and 4 only   E. 3 and 4 only

Table 1 shows that 51% of students answered question 2 correctly (choice C) but 36% of them selected an answer which included statement (1). This choice by more than one third of the students in question 2 is inconsistent with the fact that 90% of them selected $\psi(x) = \langle x|\psi\rangle$ as correct in question 1, which was immediately before question 2. Moreover, they did not realize that there was an inconsistency between their responses to questions 1 and 2. Individual interviews suggest that this inconsistency is partly due to the fact that students felt that an inner product written in Dirac notation must involve an integral. In one interview, a student who answered question 1 correctly but incorrectly claimed that only statement 1 is correct for question 2 reasoned as follows about question 2 "*Maybe [statement] (2) is correct but it just doesn't seem correct, that $\psi(x)$ should just pop out. It's giving you just a wavefunction of x and I just don't like that. I think it [inner product $\langle x|\psi\rangle$] should just give you a number.*" He correctly reasoned that the inner product is a number (with dimensions), but did not make the connection that the function $\psi(x)$ is also a number for any particular value of x. Moreover, the student did not realize the inconsistency between his responses to questions 1 and 2. Additionally, for question 2, 29% of students incorrectly selected statement (4) as correct. An interviewed student who claimed that statement (4) is correct said, "*…this inner product of p with psi is like an integral, so if you think about it… as p is the momentum operator in one dimension… that would just be $\frac{\hbar}{i}\frac{\partial}{\partial x}$ and then $\Psi(x)dx$ [gets] integrated. So I believe [statement 4]… is true.*" Further discussion suggests that the student was having difficulty in differentiating between the momentum eigenstate in dual space and the momentum operator. In particular, the student noted that the momentum eigenstate in the dual space was like the momentum operator acting on state $|\psi\rangle$. Students who had this type of difficulty generally did not realize that the momentum and position space wavefunctions are related by Fourier transform (statement (3) in question 2).

The following question was administered to investigate student difficulties with the use of the identity operator and the probability of measuring a continuous observable, e.g., position of a particle.

3. *Choose all of the following statements that are correct.*

(1) $|\psi\rangle = \int|p\rangle\langle p|\psi\rangle dp$     (2) $|\psi\rangle = \int \psi(x)|x\rangle dx$

(3) *If you measure the position of the particle in the state* $|\psi\rangle$, *the probability of finding the particle between x and x+dx is* $|\langle x|\psi\rangle|^2 dx$.

A. 1 only   B. 1 and 2 only   C. 1 and 3 only   D. 2 and 3 only   E. all of the above

Table 1 shows that only 31% of the students selected the correct answer for question 3 (Choice E) and the most common incorrect answer was choice C. Interviews suggest that while many students immediately recognized that statement (1) in question 3 is correct because a complete set of momentum eigenstates can be inserted using the identity operator, they did not realize that a similar reasoning can be used to assess the validity of statement (2) (surprisingly, this difficulty was common even if they correctly noted in question 1 that $\psi(x) = \langle x|\psi\rangle$). For example, one student who correctly noted that $\psi(x) = \langle x|\psi\rangle$ and $\psi(p) = \langle p|\psi\rangle$ in question 1 are correct and $|\psi\rangle = \int |p\rangle\langle p|\psi\rangle dx$ (which is statement (1) in question 3) is also correct said the following about statement (2) in question 3, *"Option [statement] 2 doesn't look right to me, I don't remember seeing anything like that. It's not an expansion of any sort that we have learned."* In the interviews, during the clarification phase, even after being explicitly asked to expand the state vector in terms of the identity operator written in terms of a complete set of position eigenstates, i.e., $|\psi\rangle = \int |x\rangle\langle x|\psi\rangle dx$, and asking them to make use of $\psi(x) = \langle x|\psi\rangle$ to judge the validity of statement (2) in question 3, some students still had difficulty. They did not realize that the inner product $\langle x|\psi\rangle$ can be moved around inside the integral (since it is just a number). Some students were also confused about the validity of statement (3) in question 3. For example, one interviewed student said, *"I wasn't sure if [statement] (3) was right, because when you say the probability of finding the particle somewhere, that should be a number, so it was throwing me off that there was a dx in there ...how can the probability depend on x?"*

The following question probes understanding of operators, particularly in the position representation.

4. An operator $\hat{Q}$ corresponding to a physical observable in the position representation is $Q(x)$. Choose all of the following statements that are correct.

(1) $\langle x|\hat{Q}|\psi\rangle = Q(x)\langle x|\psi\rangle$

(2) $\langle x|\hat{Q}|\psi\rangle = Q(x)\psi(x)$

(3) $\langle x|\hat{Q}|\psi\rangle = \langle \psi|\hat{Q}|x\rangle$

A. 1 only  B. 2 only  C. 1 and 2 only  D. 1 and 3 only  E. 2 and 3 only

Table 1 shows that for question 4, 54% of students chose the correct answer (Choice C) but 41% of them incorrectly claimed that either statement (1) is correct but not statement (2) or vice versa, even though in question 1, 90% of them selected $\psi(x) = \langle x|\psi\rangle$. This discrepancy again indicates that students are inconsistent in their reasoning and their ability to use Dirac notation correctly is context dependent. Additionally, 18% of students incorrectly claimed that statement (3) in question 4 is correct. An interviewed student explained his reasoning for choosing statement (3) as correct as follows, *"[statement] (3) is definitely correct, because it's [the operator $\hat{Q}$] hermitian, it's an observable."* Another interviewed student tried to use the identity operator written in terms of the eigenstates of $\hat{Q}$ and inserted it between the operator $\hat{Q}$ and |Ψ⟩, but as he worked through it mathematically he did not obtain an expression that gave him any further insight into the correctness of statements (1) or (2) and he incorrectly concluded that those statements must be incorrect. He only chose statement (3) as correct saying *"[statement] (3) is correct because I know it's always going to come out a real number for a physical observable. And the two numbers on each side will be just the complex conjugation of each one, and since it's a real number they [statements (2) and (3)] should be the same."* While the student correctly noted that the operator $\hat{Q}$ is hermitian because it corresponds to a physical observable and that the eigenvalues of $\hat{Q}$ are real, he used an incorrect reasoning that since the eigenvalues of $\hat{Q}$ are real, one can exchange the bra and ket in statement (3) without complex conjugation.

The following question probes students' understanding of quantum measurement and the measurement probabilities.

5. An operator $\hat{Q}$ corresponding to a physical observable Q has a continuous non-degenerate spectrum of eigenvalues. $|\psi_q\rangle$ are eigenstates of $\hat{Q}$ with eigenvalues $q$. At time t=0, the state of the system is $|\psi\rangle$. Choose all of the following statements that are correct.

(1) A measurement of the observable Q must return one of the eigenvalues of the operator $\hat{Q}$.

(2) If you measure Q at time t=0, the probability of obtaining an outcome between $q$ and $q+dq$ is $|\langle \psi_q|\psi\rangle|^2 dq$.

(3) If you measure Q at time t=0, the probability of obtaining an outcome between $q$ and $q+dq$ is $\left|\int_{-\infty}^{+\infty} \psi_q^*(x)\psi(x)dx\right|^2 dq$ in which $\psi_q(x)$ and $\psi(x)$ are the wavefunctions corresponding to states $|\psi_q\rangle$ and $|\psi\rangle$ respectively.

A. 1 only  B. 1 and 2 only  C. 1 and 3 only  D. 2 and 3 only  E. all of the above

Table 1 shows that 41% of students answered question 5 correctly (Choice E) but 28% of them incorrectly claimed that either statement (2) or statement (3) is correct but not both. Interviews suggest that these students were having difficulty in translating the probability density from the abstract vector space to position space. Also, some students generalized their knowledge of probability for a finite interval, which would require an integral, to include infinitesimal intervals between q and q+dq and x and x+dx, which do not require integration. For example, in one think-aloud interview, a student said that statement (2) is not correct because "*you would have to integrate between q and q+dq, it's not just that thing [ $|\langle\psi_q|\psi\rangle|^2 dq$ ], that's just some step [dq] in q space. You need to integrate over all those dq's if you want to find a probability…*" The same student had chosen statement (3) in question 3 as a correct statement (which is a special case of statement (2) in question 5). In the clarification phase, when asked why there was no integral in statement (3) in question 3, he reflected on the inconsistency that was pointed out explicitly and changed his answer for question 3 which was correct to make it consistent with his answer for question 5 saying "*I do feel [now] that there should be an integral sign, because otherwise it's just a dx of your probability, and so based on just the notation of it I would say that's incorrect.*"

## SUMMARY AND FUTURE OUTLOOK

We find that many graduate students in physics struggle with Dirac notation even after instruction in graduate level quantum mechanics. In particular, many students were inconsistent in their use of Dirac notation across questions both in a written test and interviews. For example, one difficulty many graduate students have is in correctly identifying the position space and momentum space wavefunctions written in Dirac notation in different contexts. We find that graduate students who correctly identify that $|\psi\rangle = \int |x\rangle\langle x|\psi\rangle dx$ and $\psi(x) = \langle x|\psi\rangle$ are both correct may not recognize that $|\psi\rangle = \int \psi(x)|x\rangle dx$ is also correct, and this latter relation is obtained by combining the other two identities that the student was able to identify correctly. Some of the students also had difficulty in distinguishing between a state and an operator written in Dirac notation (e.g., they had difficulty in distinguishing a momentum eigenstate $|p\rangle$ in the dual space and the momentum operator). Students also struggled to recognize the probability of obtaining a particular outcome in Dirac notation, even though they correctly identified the same probability in the position representation. Pertaining to this issue, one common difficulty revealed in the interviews was related to confusion about $|\langle\psi_q|\psi\rangle|^2$ or $|\langle x|\psi\rangle|^2$ which give the probability densities for measuring an eigenvalue q or x for an observable Q or x, respectively, in a state $|\psi\rangle$. These students often incorrectly claimed that an expression for the probability of measuring an observable in an infinitesimal interval must involve integration over q or x even in the Dirac notation (e.g., they claimed that there should be an integral in statement (2) in question 5). Some students also incorrectly claimed that one can always exchange the bra and ket states in the Dirac notation if the operator sandwiched between them was a hermitian operator corresponding to an observable. While some of them correctly reasoned that the eigenvalues of a hermitian operator are real, they erroneously concluded that this implies that one can exchange the bra and ket states without complex conjugation if the scalar product involves sandwiching a hermitian operator.

Based upon the research on student difficulties, we have been developing and assessing research-based learning tools to help students develop a good grasp of Dirac notation in quantum mechanics. These research-based learning tools include the Quantum Interactive Learning Tutorial (QuILT) [7-9], reflective problems which complement textbook problems, and concept tests on Dirac notation using an approach similar to that popularized by Mazur for introductory courses [10]. The QuILT employs a guided inquiry-based approach and is designed to help students build a good knowledge structure. The instructors can use the QuILT as an in-class tutorial and students can be asked to work in small groups of two or three on the QuILT and make sense of Dirac notation in various contexts. The QuILT can be used by underprepared graduate students as a self-study tool. Reflective problems can be used as a homework supplement in undergraduate courses. The concept tests can be integrated with lectures and students can take advantage of their peers' expertise and learn from each other [10].

## ACKNOWLEDGEMENTS

We thank the National Science Foundation for awards PHY-0968891 and PHY-1202909.